\documentclass[12pt,thmsa]{article}
\usepackage{amssymb}

\usepackage{sw20aip}

\input tcilatex

\begin{document}

\title{Path integral for relativistic oscillators :model \\
of the Klein-Gordon particle in AdS space}
\author{M. T. Chefrour \\
%EndAName
D\'epartement de Physique, Facult\'e des Sciences,\\
Universit\'e Badji Mokhtar, Annaba, Algeria. \and F. Benamira and L. Guechi 
\\
%EndAName
Laboratoire de Physique Th\'{e}orique,\\
D\'{e}partement de Physique, Facult\'{e} des Sciences,\\
Universit\'e Mentouri, Route d'Ain El Bey, \\
Constantine, Algeria.}
\date{\today }
\maketitle

\begin{abstract}
Explicit path integration is carried out for the Green's functions of
special relativistic harmonic oscillators in (1+1)- and (3+1)-dimensional
Minkowski space-time modeled by a Klein-Gordon particle in the universal
covering space-time of the anti-de Sitter static space-time. The energy
spectrum together with the normalized wave functions are obtained. In the
non-relativistic limit, the bound states of the one- and three-dimensional
ordinary oscillators are regained.


typescript using Latex ( version 2.0)
\end{abstract}

\section{Introduction}

Relativistic problems that can be solved exactly by the use of the path
integral approach are very limited especially for two reasons:

(1) For a relativistic particle with spin, the propagator cannot be
described by a simple path integral based on any reasonable action. The fact
that the spin has no classical origin makes it difficult to propose for it
continuous paths\cite{FeyHib} .

(2) If the particles interact with each other or with an external potential,
they can produce quantum effects which cannot be described by path
fluctuations alone. These effects can be handled by perturbation theory in
the framework of the quantum field theory\cite{Klei1} .

However, in recent years, there have been a few successful examples where
the difficulty which concerns the spin has been shaped. The Dirac propagator
for a free particle\cite{BarDur} has been derived in the framework of a
model where the spin is classically described by internal variables. The
path integral treatments of the Dirac-Coulomb problem\cite{KayIno} and a
Dirac electron in a one-dimensional Coulomb potential on the half-line and
in the presence of an external superstrong magnetic field\cite{BerCarLim}
have been obtained via the Biedenharn transformation\cite{Bied} . The
electron in the presence of a constant magnetic field\cite{PapDev} and the
problem of charged particles in interaction with an electromagnetic plane
wave alone \cite{BouCheGueHam} or plus a parallel magnetic field\cite
{BenCheGueHam} have been studied by introducing a fifth parameter in order
to bring the problem into a non-relativistic form. The relativistic spinless
Coulomb system\cite{Klei2} and the Klein-Gordon particle in vector plus
scalar Hulth\'{e}n-type potentials\cite{CheGueLecHamMes} have also been
solved by path integration.

Recently, from various aspects (\cite
{DulVan,MosSze,BenMarRomNun,AldBisNav,NavNav} and references therein), there
has been renewed interest for the relativistic harmonic oscillators because
of a crucial point. Indeed, a simple replacement of the coordinates and
generalized momenta in the corresponding classical Hamiltonian by their
quantum mechanical counterparts is, in general, not correct since the
ambiguity resulting from ordering the operators must be resolved. To
parameterize the operator ordering ambiguity of the position- and the
momentum-operators, we show that it is necessary to introduce two parameters 
$\alpha $ and $\beta $ which can not be freely chosen. The problem of the
quantum relativistic oscillators represented by quantum free relativistic
particles on the universal covering space-time of the anti- de Sitter static
space-time (CAdS) is a model characterized by a constraint on these
parameters. This model is called '' special quantum relativistic oscillators
'' in the sense that $\alpha $ and $\beta $ are chosen to adjust the
non-relativistic limit and to preserve the reality of the energy spectrum of
the physical system.

To our knowledge, there is no path integral discussion for the quantum
relativistic harmonic oscillators. The purpose of the present paper is to
fill this gap. The treatment will be restricted to spinless systems.

Our study is organized in the following way: in sec. II, we construct the
path integral associated with the $(1+1)$-dimensional special relativistic
harmonic oscillator. The Green's function is derived in closed form, from
which we obtain the energy spectrum and the normalized wave functions. In
sec. III, we extend the discussion to the $(3+1)$-dimensional case. The
radial Green's function is also given in closed form. The energy levels and
the normalized wave functions are then deduced. The section IV will be a
conclusion.

\section{The (1+1)-dimensional special relativistic oscillator}

The relativistic harmonic oscillator interaction in $(1+1)$ Minkowski
space-time is equivalent to a free relativistic particle in the universal
covering space-time of the anti-de Sitter space-time (CAdS). For a static
form of the anti-de Sitter space-time metric, the line element is given by

\begin{equation}
ds^2=\Lambda (x)c^2dt^2-\frac 1{\Lambda (x)}dx^2,  \label{a.1}
\end{equation}
where 
\begin{equation}
\Lambda (x)=1+\frac{\omega ^2}{c^2}x^2.  \label{a.2}
\end{equation}
Classical mechanics is described in this space by the classical Lagrangian
and Hamiltonian, respectively:

\begin{equation}
L=-Mc\sqrt{1-\frac 1{\Lambda (x)}\frac{v^2}{c^2}+\frac{\omega ^2}{c^2}x^2},
\label{a.3}
\end{equation}

\begin{equation}
H^2=M^2c^4+p^2c^2+M^2\omega ^2c^2x^2+2\omega ^2x^2p^2+\frac{\omega ^4}{c^2}%
x^4p^2.  \label{a.4}
\end{equation}

If we proceed by adopting the substitutions $H\rightarrow \widehat{P}%
_{0}=i\hbar \frac{\partial }{\partial t},$ $p\rightarrow \widehat{P}_{x}=%
\frac{\hbar }{i}\frac{\partial }{\partial x},$ $x\rightarrow \widehat{x}$,
there is an ambiguity which results from ordering \ the operators in the
quantum mechanical counterpart of (\ref{a.4}). Since, there exist different
ways to put the terms $x^{4}p^{2}$and $x^{2}p^{2}$into symmetrically ordered
forms, we can construct a number of Hermitian mechanical quantum
counterparts of (\ref{a.4}). In order to avoid this ambiguity, we write all
the Hermitian forms for each term as a linear combination. Whence, after
calculation of all the commutators, we find the following replacements 
\begin{equation}
\left\{ 
\begin{array}{c}
x^{4}p^{2}\rightarrow -\hbar ^{2}\left( x^{4}\frac{d^{2}}{dx^{2}}+4x^{3}%
\frac{d}{dx}+\alpha x^{2}\right) , \\ 
x^{2}p^{2}\rightarrow -\hbar ^{2}\left( x^{2}\frac{d^{2}}{dx^{2}}+2x\frac{d}{%
dx}+\beta \right) ,
\end{array}
\right.  \label{a.5}
\end{equation}
where the parameters $\alpha $ and $\beta $ will be fixed farther.

The Green's function $G(x^{\prime \prime },x^{\prime })$ that we consider
obeys the Klein-Gordon equation

\begin{equation}
\left\{ \Box +\kappa \Lambda (x)+(1-\alpha +2\beta )\frac{\omega ^4}{c^4}x^2+%
\frac{\omega ^2}{c^2}\right\} G(x^{\prime \prime },x^{\prime })=-\frac
1{\hbar ^2c^2}\delta \left( x^{\prime \prime }-x^{\prime }\right) .
\label{a.6}
\end{equation}
where

\begin{equation}
\Box =\frac{1}{c^{2}}\frac{\partial ^{2}}{\partial t^{2}}-\Lambda (x)\frac{%
\partial ^{2}}{\partial x^{2}}\Lambda (x),  \label{a.7}
\end{equation}
and

\begin{equation}
\kappa =\left( \frac{Mc}{\hbar }\right) ^{2}+\left( 1-2\beta \right) \frac{%
\omega ^{2}}{c^{2}}.  \label{a.8}
\end{equation}
Note that choosing to work with the symmetrically ordered form (\ref{a.7})
of the D'Alembertian operator, the quantization of the original problem will
not be modified.

By using the Schwinger's integral representation \cite{Schwinger} , the
solution of the differential equation (\ref{a.6}) can be written as follows:

\begin{equation}
G(x^{\prime \prime },x^{\prime })=\frac 1{2i\hbar c^2}\int_0^\infty d\lambda
\left\langle x^{\prime \prime },t^{\prime \prime }\right| \exp \left[ \frac
i\hbar \widehat{H}\lambda \right] \left| x^{\prime },t^{\prime
}\right\rangle ,  \label{a.9}
\end{equation}
where the integrand $\left\langle x^{\prime \prime },t^{\prime \prime
}\right| \exp \left[ \frac i\hbar \widehat{H}\lambda \right] \left|
x^{\prime },t^{\prime }\right\rangle $ is similar to the propagator of a
quantum system evolving in $\lambda $ time from $(x^{\prime },t^{\prime })$
to $(x^{\prime \prime },t^{\prime \prime })$ with the effective Hamiltonian,

\begin{equation}
\widehat{H}=\frac 12\left[ \!-\Lambda (x)\widehat{P}_x^2\Lambda (x)+\frac{%
\widehat{P}_0^2}{c^2}-\hbar ^2\kappa \Lambda (x)-\hbar ^2(1\!-\!\alpha
+\!2\beta )\frac{\omega ^4}{c^4}x^2-\frac{\hbar ^2\omega ^2}{c^2}\right] .
\label{a.10}
\end{equation}
The integrand in Eq. (\ref{a.9}) may be written as the path integral \cite
{BouCheGueHam,BenCheGueHam,CheGueLecHamMes,Feyn,Schulm}

\begin{eqnarray}
P(x^{\prime \prime },t^{\prime \prime },x^{\prime },t^{\prime };\lambda )
&=&\left\langle x^{\prime \prime },t^{\prime \prime }\right| \exp \left[
\frac i\hbar \widehat{H}\lambda \right] \left| x^{\prime },t^{\prime
}\right\rangle  \nonumber \\
&=&\stackunder{N\rightarrow \infty }{\lim }\!\int \!\stackrel{N}{\stackunder{%
n=1}{\prod }}\!dx_ndt_n\!\stackrel{N+1}{\stackunder{n=1}{\prod }}\!\frac{%
d(P_x)_n}{2\pi \hbar }\frac{d(P_0)_n}{2\pi \hbar }\exp \left\{ \!\frac
i\hbar \stackunder{n=1}{\!\stackrel{N+1}{\sum }\!}A_1^\varepsilon \right\} ,
\nonumber \\
&&  \label{a.11}
\end{eqnarray}

with the short-time action 
\begin{eqnarray}
A_{1}^{\varepsilon } &=&(P_{0})_{n}\triangle t_{n}-(P_{x})_{n}\triangle
x_{n}+\frac{\varepsilon }{2}\left( \frac{(P_{0})_{n}^{2}}{c^{2}}-\Lambda
(x_{n})\Lambda (x_{n-1})(P_{x})_{n}^{2}\right.  \nonumber \\
&&\left. -\hbar ^{2}\frac{\omega ^{2}}{c^{2}}-\hbar ^{2}\kappa \Lambda
(x_{n})-\hbar ^{2}(1-\alpha +2\beta )\frac{\omega ^{4}}{c^{4}}%
x_{n}^{2}\right) ,  \label{a.12}
\end{eqnarray}
where 
\begin{equation}
\varepsilon =\frac{\lambda }{N+1}=s_{n}-s_{n-1},  \label{a.13}
\end{equation}
and $s\in \left[ 0,\lambda \right] $ is a new time-like variable.

Let us first notice that the integrations on the variables $t_{n}$ give $N$
Dirac distributions $\delta \left( (P_{0})_{n}-(P_{0})_{n+1}\right) .$
Thereafter the integrations on $(P_{0})_{n}$ give $%
(P_{0})_{1}=(P_{0})_{2}=...=(P_{0})_{N+1}=E.$ The propagator (\ref{a.11})
then becomes

\begin{equation}
P(x^{\prime \prime },t^{\prime \prime },x^{\prime },t^{\prime };\lambda
)=\int_{-\infty }^{+\infty }\frac{dE}{2\pi \hbar }\exp \left[ -\frac i\hbar
E(t^{\prime \prime }-t^{\prime })\right] P_E(x^{\prime \prime },x^{\prime
};\lambda ),  \label{a.14}
\end{equation}
where the kernel $P_E(x^{\prime \prime },x^{\prime };\lambda )$ is given by

\begin{equation}
P_{E}(x^{\prime \prime },x^{\prime };\lambda )=\stackunder{N\rightarrow
\infty }{\lim }\int \stackrel{N}{\stackunder{n=1}{\prod }}dx_{n}\stackrel{N+1%
}{\stackunder{n=1}{\prod }}\frac{d(P_{x})_{n}}{2\pi \hbar }\exp \left[ \frac{%
i}{\hbar }\stackunder{n=1}{\stackrel{N+1}{\sum }}A_{2}^{\varepsilon }\right]
,  \label{a.15}
\end{equation}
with the short-time action

\begin{eqnarray}
A_2^\varepsilon &=&-(P_x)_n\triangle x_n+\frac \varepsilon 2\left[ -\Lambda
(x_n)\Lambda (x_{n-1})(P_x)_n^2+\frac{\hbar ^2\omega ^2}{c^2}\left( \frac{E^2%
}{\hbar ^2\omega ^2}-1\right) \right.  \nonumber \\
&&\left. -\hbar ^2\kappa \Lambda (x_n)-\hbar ^2(1-\alpha +2\beta )\frac{%
\omega ^4}{c^4}x_n^2\right] .  \label{a.16}
\end{eqnarray}
Note that (\ref{a.14}) is invariant under the change $E\rightarrow -E.$

Then, by integrating with respect to the variables $(P_x)_n$, we get

\begin{eqnarray}
P_E(x^{\prime \prime },x^{\prime };\lambda ) &=&\frac 1{\sqrt{\Lambda
(x^{\prime })\Lambda (x^{\prime \prime })}}\stackunder{N\rightarrow \infty }{%
\lim }\stackrel{N+1}{\stackunder{n=1}{\prod }}\left[ \frac 1{2i\pi \hbar
\varepsilon }\right] ^{\frac 12}  \nonumber \\
&&\times \stackrel{N}{\stackunder{n=1}{\prod }}\left[ \int \frac{dx_n}{%
\Lambda (x_n)}\right] \exp \left[ \frac i\hbar \stackunder{n=1}{\stackrel{N+1%
}{\sum }}A_3^\varepsilon \right] ,  \label{a.17}
\end{eqnarray}
with the short-time action in configuration space

\begin{eqnarray}
A_3^\varepsilon &=&\frac{\triangle x_n^2}{2\varepsilon \Lambda (x_n)\Lambda
(x_{n-1})}+\frac \varepsilon 2\hbar ^2\left[ \left( \frac{E^2}{\hbar
^2\omega ^2}-1\right) \frac{\omega ^2}{c^2}-\kappa \Lambda (x_n)\right. 
\nonumber \\
&&\left. -(1-\alpha +2\beta )\frac{\omega ^4}{c^4}x_n^2\right] .
\label{a.18}
\end{eqnarray}

Substituting (\ref{a.14}) into (\ref{a.9}), we can rewrite (\ref{a.9}) in
the form:

\begin{equation}
G(x^{\prime \prime },x^{\prime })=\int_{-\infty }^{+\infty }\frac{dE}{2\pi
\hbar }\exp \left[ -\frac i\hbar E(t^{\prime \prime }-t^{\prime })\right]
G_E(x^{\prime \prime },x^{\prime }),  \label{a.19}
\end{equation}
with 
\begin{equation}
G_E(x^{\prime \prime },x^{\prime })=\frac 1{2i\hbar c^2}\int_0^\infty
d\lambda P_E(x^{\prime \prime },x^{\prime };\lambda ).  \label{a.20}
\end{equation}

If we now introduce a new variable $u_{n}$ together with a rescaling of time 
\cite{DurKlei} from $\varepsilon $ to $\sigma _{n\text{ }}$given by

\begin{equation}
\left\{ 
\begin{array}{c}
x_n=\frac c\omega \sinh u_n, \\ 
\\ 
\varepsilon =\sigma _n\frac{c^2}{\omega ^2}\frac 1{\cosh u_n\cosh u_{n-1}},
\end{array}
\right.  \label{a.21}
\end{equation}
and incorporate the constraint

\begin{equation}
\lambda =\frac{c^2}{\omega ^2}\int_0^S\frac{ds}{\cosh ^2u},  \label{a.22}
\end{equation}
by using the identity

\begin{equation}
\frac{c^{2}/\omega ^{2}}{\cosh u^{\prime \prime }\cosh u^{\prime }}%
\int_{0}^{\infty }dS\delta \left( \lambda -\frac{c^{2}}{\omega ^{2}}%
\int_{0}^{S}\frac{ds}{\cosh ^{2}u}\right) =1,  \label{a.23}
\end{equation}
the path integral (\ref{a.20}) can be written as:

\begin{equation}
G_E(x^{\prime \prime },x^{\prime })=\frac 1{2i\hbar \omega c\left( \cosh
u^{\prime \prime }\cosh u^{\prime }\right) ^{\frac 32}}\int_0^\infty
dSP(u^{\prime \prime },u^{\prime };S),  \label{a.24}
\end{equation}
where

\begin{eqnarray}
P(u^{\prime \prime },u^{\prime };S) &=&\stackunder{N\rightarrow \infty }{%
\lim }\int \stackrel{N+1}{\stackunder{n=1}{\prod }}\left[ \frac{1}{2i\pi
\hbar \sigma _{n}}\right] ^{\frac{1}{2}}\stackrel{N}{\stackunder{n=1}{\prod }%
}du_{n}\exp \left\{ \frac{i}{\hbar }\stackunder{n=1}{\stackrel{N+1}{\sum }}%
\left[ \frac{\triangle u_{n}^{2}}{2\sigma _{n}}\right. \right.  \nonumber \\
&&\ +\frac{\triangle u_{n}^{4}}{8\sigma _{n}}\left( \frac{1}{3}-\frac{1}{%
\cosh ^{2}\widetilde{u}_{n}}\right) -\frac{\hbar ^{2}}{2}\left( \frac{c^{2}}{%
\omega ^{2}}\kappa -\frac{E^{2}/\hbar ^{2}\omega ^{2}-1}{\cosh ^{2}%
\widetilde{u}_{n}}\right.  \nonumber \\
&&\ \left. \left. \left. +(1-\alpha +2\beta )\tanh ^{2}\widetilde{u}%
_{n}\right) \sigma _{n}\right] \right\} .  \label{a.25}
\end{eqnarray}
Here, we have used the usual abbreviations $\triangle u_{n}=u_{n}-u_{n-1},$ $%
\widetilde{u}_{n}=\frac{u_{n}+u_{n-1}}{2},$ $u^{\prime }=u(0)$ and $%
u^{\prime \prime }=u(S).$ Note that the term in $(\triangle u_{n})^{4}$
contributes significantly to the path integral. It can be estimated by using
the formula \cite{Dewitt}

\begin{equation}
\int_{-\infty }^{+\infty }\exp (-\alpha _{1}x^{2}+\alpha
_{2}x^{4})dx=\int_{-\infty }^{+\infty }\exp \left( -\alpha _{1}x^{2}+\frac{%
3\alpha _{2}}{4\alpha _{1}^{2}}\right) dx,  \label{a.26}
\end{equation}
valid for $\left| \alpha _{1}\right| $ large and Re$(\alpha _{1})>0.$ This
leads to

\begin{eqnarray}
P(u^{\prime \prime },u^{\prime };S) &=&\int Du(s)\exp \left\{ \frac{i}{\hbar 
}\int_{0}^{S}\left[ \frac{\stackrel{.}{u}^{2}}{2}-\frac{\hbar ^{2}}{2}\left( 
\frac{c^{2}}{\omega ^{2}}\kappa +\frac{1}{4}\right) \right. \right. 
\nonumber \\
&&\left. \left. +\frac{\hbar ^{2}}{2}\frac{E^{2}/\hbar ^{2}\omega ^{2}-1/4}{%
\cosh ^{2}u}-\frac{\hbar ^{2}}{2}(1-\alpha +2\beta )\tanh ^{2}u\right]
ds\right\} .  \nonumber \\
&&  \label{a.27}
\end{eqnarray}
By noting that $\tanh ^{2}u=1-\frac{1}{\cosh ^{2}u},$ this last path
integral is identical in form with that of the symmetric Rosen-Morse
potential \cite{RosMor} which has been studied recently \cite
{JunIno,BohJun,Grosc,KleiMus,CheGueHamLec} , but in order to obtain the
equivalent to the Klein-Gordon equation in the AdS space-time we impose a
restriction on the parameters $\alpha $ and $\beta $ defined by the
following two equations:

\begin{equation}
1-\alpha +2\beta =0,  \label{a.28}
\end{equation}

\begin{equation}
(1-2\beta )\frac{\omega ^2}{c^2}=\xi R,  \label{a.29}
\end{equation}
where $R=-2\frac{\omega ^2}{c^2}$ is the scalar curvature and $\xi $ is a
numerical factor. Whence it follows that

\begin{equation}
\alpha =2\xi +2\text{\qquad and\qquad }\beta =\xi +\frac{1}{2}.  \label{a.30}
\end{equation}
In this case, the propagator (\ref{a.27}) reduces to 
\begin{eqnarray}
P(u^{\prime \prime },u^{\prime };S) &=&\int Du(s)\exp \left\{ \frac{i}{\hbar 
}\int_{0}^{S}\left[ \frac{\stackrel{.}{u}^{2}}{2}-\frac{\hbar ^{2}}{2}\left(
\left( \frac{Mc^{2}}{\hbar \omega }\right) ^{2}-2\xi +\frac{1}{4}\right)
\right. \right.  \nonumber \\
&&\left. \left. +\frac{\hbar ^{2}}{2}\frac{E^{2}/\hbar ^{2}\omega ^{2}-1/4}{%
\cosh ^{2}u}\right] ds\right\} ,  \label{a.31}
\end{eqnarray}
which is likewise the propagator relative to a symmetric Rosen-Morse
potential. The Green's function associated with this potential has been
evaluated through various techniques of path integration \cite
{JunIno,BohJun,Grosc,KleiMus,CheGueHamLec} . The result is

\begin{eqnarray}
G(u^{\prime \prime },u^{\prime };E) &=&\int_{0}^{+\infty }dSP(u^{\prime
\prime },u^{\prime };S)  \nonumber \\
&=&-\frac{i}{\hbar }\Gamma (\gamma -l_{E})\Gamma (1+l_{E}+\gamma
)P_{l_{E}}^{-\gamma }(\tanh u^{\prime \prime })P_{l_{E}}^{-\gamma }(-\tanh
u^{\prime }),  \nonumber \\
&&  \label{a.32}
\end{eqnarray}
where $P_{l_{E}}^{-\gamma }(\tanh u)$ is the associated Legendre function
with

\begin{equation}
l_E=-\frac 12+\frac E{\hbar \omega },  \label{a.33}
\end{equation}
and 
\begin{equation}
\gamma =\pm \frac 12\sqrt{1+4N^2-8\xi },\quad N=\frac{Mc^2}{\hbar \omega }.
\label{a.34}
\end{equation}

If we take into account Eqs. (\ref{a.30}), insert (\ref{a.32}) into (\ref
{a.24}), and remember the first equation of the transformation (\ref{a.21}),
we obtain the Green's function for the one-dimensional special relativistic
harmonic oscillator under consideration

\begin{eqnarray}
G_E(x^{\prime \prime },x^{\prime }) &=&-\frac{\Gamma (\gamma -l_E)\Gamma
(1+l_E+\gamma )}{2\hbar ^2\omega c}\left[ \left( 1+\frac{\omega ^2}{c^2}%
x^{\prime \prime }{}^2\right) \left( 1+\frac{\omega ^2}{c^2}x^{\prime
2}\right) \right] ^{-\frac 34}  \nonumber \\
&&\times P_{l_E}^{-\gamma }\left( \frac{\frac \omega cx^{\prime \prime }}{%
\sqrt{1+\frac{\omega ^2}{c^2}x^{\prime \prime 2}}}\right) P_{l_E}^{-\gamma
}\left( -\frac{\frac \omega cx^{\prime }}{\sqrt{1+\frac{\omega ^2}{c^2}%
x^{\prime 2}}}\right) .  \label{a.35}
\end{eqnarray}

The poles of the Green's function yield the discrete energy spectrum. These
are just the poles of $\Gamma (\gamma -l_E)$ which occur when $\gamma
-l_E=-n $ for $n=0,1,2,...$. They are given through the equations

\begin{equation}
\frac{1}{2}-\frac{E}{\hbar \omega }\pm \frac{1}{2}\sqrt{1+4N^{2}-8\xi }=-n.
\label{a.36}
\end{equation}
So, algebraically we obtain two distinct sets of energy levels according to
the positive and negative signs of the parameter $\gamma $. But we have to
check whether the corresponding wave functions, which will be expressed in
terms of the Legendre functions of the first kind $P_{l_{E}}^{-\gamma }(y),$
satisfy the boundary conditions for $y=\frac{\omega }{c}x/\sqrt{1+\frac{%
\omega ^{2}}{c^{2}}x^{2}}\rightarrow \pm 1.$ By inspecting their asymptotic
behaviors\cite{Erdelyi}

\begin{equation}
P_{l_{E}}^{-\gamma }(y)\stackunder{y\rightarrow 1}{\simeq }\frac{(1-y)^{%
\frac{\gamma }{2}}}{2^{\frac{\gamma }{2}}\Gamma (1+\gamma )};\qquad \gamma
\neq 0,-1,-2,-3,...,  \label{a.37}
\end{equation}
\begin{equation}
P_{l_{E}}^{-\gamma }(y)\stackunder{y\rightarrow -1}{\simeq }\left\{ 
\begin{array}{c}
-\frac{\Gamma (-\gamma )}{2^{\frac{\gamma }{2}}\pi }\sin (l_{E}\pi )(1+y)^{%
\frac{\gamma }{2}}\qquad \text{for\quad Re}(\gamma )<0, \\ 
\\ 
\frac{2^{\frac{\gamma }{2}}\Gamma (\gamma )}{\Gamma (1+l_{E}+\gamma )\Gamma
(\gamma -l_{E})}(1+y)^{-\frac{\gamma }{2}}\qquad \text{for\quad Re}(\gamma
)>0,
\end{array}
\right.  \label{a.38}
\end{equation}
we see that $P_{l_{E}}^{-\gamma }(y)$ diverges if Re$(\gamma )<0$.
Therefore, we must choose the positive sign of $\gamma $ and hence the
energy eigenvalues are

\begin{equation}
E_{n}=\left( n+\frac{1}{2}+\frac{1}{2}\sqrt{1+4N^{2}-8\xi }\right) \hbar
\omega .  \label{a.39}
\end{equation}
On the other hand, the reality of the parameter $\gamma $ implies the
following range of the numerical factor $\xi <\frac{1}{8}\left(
1+4N^{2}\right) $.

In the limit $c\rightarrow \infty $, the energy spectrum approaches

\begin{equation}
E_n^{NR}+Mc^2=\hbar \omega \left( n+\frac 12\right) +Mc^2.  \label{a.40}
\end{equation}
The first term gives the energy levels in the non-relativistic case and the
second term is the rest energy of the harmonic oscillator.

The corresponding energy eigenfunctions can be found by approximation near
the poles $\gamma -l_E\approx -n$ :

\begin{equation}
\Gamma \left( \gamma -l_{E}\right) \approx \frac{(-1)^{n}}{n!}\frac{1}{%
\gamma -l_{E}+n}=\frac{(-1)^{n+1}}{n!}\frac{2\left( n+\gamma +\frac{1}{2}%
\right) \hbar ^{2}\omega ^{2}}{E^{2}-E_{n}^{2}}.  \label{a.41}
\end{equation}
Using this behavior and the known property of the symmetry of the associated
Legendre functions under spatial reflection, $x\rightarrow -x$, we get the
contribution of the bound states to the spectral representation of the
Green's function as

\begin{eqnarray}
G_{E}(x^{\prime \prime },x^{\prime }) &=&\stackunder{n=0}{\stackrel{\infty }{%
\sum }}\frac{\omega }{c}\frac{n+\gamma +\frac{1}{2}}{n!\left(
E^{2}-E_{n}^{2}\right) }\Gamma (2\gamma +n+1)  \nonumber \\
&&\times \left( 1+\frac{\omega ^{2}}{c^{2}}x^{\prime \prime 2}\right) ^{-%
\frac{3}{4}}\left( 1+\frac{\omega ^{2}}{c^{2}}x^{\prime 2}\right) ^{-\frac{3%
}{4}}  \nonumber \\
&&\times P_{n+\gamma }^{-\gamma }\left( \frac{\frac{\omega }{c}x^{\prime
\prime }}{\sqrt{1+\frac{\omega ^{2}}{c^{2}}x^{\prime \prime 2}}}\right)
P_{n+\gamma }^{-\gamma }\left( \frac{\frac{\omega }{c}x^{\prime }}{\sqrt{1+%
\frac{\omega ^{2}}{c^{2}}x^{\prime 2}}}\right)  \nonumber \\
\ &=&\stackunder{n=0}{\stackrel{\infty }{\sum }}\frac{\Psi _{n}^{\gamma
}(x^{\prime \prime })\Psi _{n}^{\gamma \ast }(x^{\prime })}{E^{2}-E_{n}^{2}}.
\label{a.42}
\end{eqnarray}
The properly normalized wave functions are thus

\begin{eqnarray}
\Psi _{n}^{\gamma }(x) &=&\left[ \frac{\omega }{c}\frac{n+\gamma +\frac{1}{2}%
}{n!}\Gamma (2\gamma +n+1)\right] ^{\frac{1}{2}}\left( 1+\frac{\omega ^{2}}{%
c^{2}}x^{2}\right) ^{-\frac{3}{4}}  \nonumber \\
&&\times P_{n+\gamma }^{-\gamma }\left( \frac{\frac{\omega }{c}x}{\sqrt{1+%
\frac{\omega ^{2}}{c^{2}}x^{2}}}\right) .  \label{a.43}
\end{eqnarray}
Taking into account the relation between the Gegenbauer polynomials and the
associated Legendre functions ( see formula (8.936) p. 1031 in Ref. \cite
{GradRyz} )

\begin{equation}
C_{n}^{\lambda }(t)=\frac{\Gamma (2\lambda +n)\Gamma (\lambda +\frac{1}{2})}{%
\Gamma (2\lambda )\Gamma (n+1)}\left[ \frac{1}{4}\left( t^{2}-1\right) %
\right] ^{\frac{1}{4}-\frac{\lambda }{2}}P_{\lambda +n-\frac{1}{2}}^{\frac{1%
}{2}-\lambda }(t),  \label{a.44}
\end{equation}
and using the doubling formula (see Eq. (8.335.1), p. 938 in Ref. \cite
{GradRyz} )

\begin{equation}
\Gamma (2x)=\frac{2^{2x-1}}{\sqrt{\pi }}\Gamma (x)\Gamma \left( x+\frac{1}{2}%
\right) ,  \label{a.45}
\end{equation}
we can also express (\ref{a.43}) in the form:

\begin{eqnarray}
\Psi _{n}^{\gamma }(x) &=&\left[ \frac{\omega }{c}\frac{\left( \gamma +n+%
\frac{1}{2}\right) n!}{\Gamma (2\gamma +n+1)}\right] ^{\frac{1}{2}%
}(2i)^{\gamma }\Gamma \left( \gamma +\frac{1}{2}\right) \left( 1+\frac{%
\omega ^{2}}{c^{2}}x^{2}\right) ^{-\frac{1}{2}\left( \gamma +\frac{3}{2}%
\right) }  \nonumber \\
&&\times C_{n}^{\gamma +\frac{1}{2}}\left( \frac{\frac{\omega }{c}x}{\sqrt{1+%
\frac{\omega ^{2}}{c^{2}}x^{2}}}\right) .  \label{a.46}
\end{eqnarray}

In the limit $c\rightarrow \infty $, $\gamma \rightarrow N=\frac{Mc^2}{\hbar
\omega }$ and with the help of the formula ( see Eq. (8.328.1), p. 937 in
Ref. \cite{GradRyz} )

\begin{equation}
\stackunder{z\rightarrow \infty }{\lim }\frac{\Gamma (z+a)}{\Gamma (z)}%
e^{-a\ln z}=1,  \label{a.47}
\end{equation}
we see that

\begin{eqnarray}
&&\stackunder{c\rightarrow \infty }{\lim }\gamma ^{\frac n2}\left[ \frac
\omega c\frac{\left( \gamma +n+\frac 12\right) n!}{\Gamma (2\gamma +n+1)}%
\right] ^{\frac 12}(2)^\gamma \Gamma \left( \gamma +\frac 12\right) 
\nonumber \\
&=&\stackunder{c\rightarrow \infty }{\lim }\left[ \frac \omega {c\sqrt{\pi }}%
\frac{n!}{2^n}\right] ^{\frac 12}\left[ \frac{\Gamma \left( \gamma +\frac
12\right) }{\Gamma (\gamma )}\right] ^{\frac 12}=\left( \frac{M\omega }{\pi
\hbar }\right) ^{\frac 14}\sqrt{\frac{n!}{2^n}}.  \label{a.48}
\end{eqnarray}
By the use of the limit relation ( see Eq. (8.936.5), p. 1031 in Ref. \cite
{GradRyz} )

\begin{equation}
\stackunder{\lambda \rightarrow \infty }{\lim }\lambda ^{-\frac
n2}C_n^{\frac \lambda 2}\left( t\sqrt{\frac 2\lambda }\right) =\frac{%
2^{-\frac n2}}{n!}H_n(t),  \label{a.49}
\end{equation}
the wave functions of the harmonic oscillator in the non-relativistic
approximation are naturally regained

\begin{equation}
\stackunder{c\rightarrow \infty }{\lim }\Psi _{n}^{\gamma }(x)=\left( \frac{%
M\omega }{\pi \hbar }\right) ^{\frac{1}{4}}\frac{1}{\sqrt{2^{n}n!}}e^{-\frac{%
M\omega }{2\hbar }x^{2}}H_{n}\left( \sqrt{\frac{M\omega }{\hbar }}x\right) ,
\label{a.50}
\end{equation}
where $H_{n}\left( \sqrt{\frac{M\omega }{\hbar }}x\right) $ is the Hermite
polynomial of n$^{\text{th}}$ order.

\section{The (3+1)-dimensional special relativistic oscillator}

The special relativistic harmonic oscillator in $(3+1)$ Minkowski space-time
is simulated in the universal covering space-time (CAdS) of the anti- de
Sitter space-time with a negative curvature $R=-12\frac{\omega ^{2}}{c^{2}}$
and a static metric of the form:

\begin{equation}
ds^2=\Lambda (r)c^2dt^2-\frac 1{\Lambda (r)}dr^2-r^2\left( d\theta ^2+\sin
^2\theta d\phi ^2\right) ,  \label{a.51}
\end{equation}
where

\begin{equation}
\Lambda (r)=1+\frac{\omega ^2}{c^2}r^2  \label{a.52}
\end{equation}
is chosen in order to impose the non-relativistic limit.

The Lagrangian reads as:

\begin{equation}
L=-Mc\sqrt{\Lambda (r)-\frac{v^2}{c^2}+\frac{\omega ^2}{c^4}\frac{(\stackrel{%
\rightarrow }{r}\stackrel{\rightarrow }{v})^2}{\Lambda (r)}}  \label{a.53}
\end{equation}
and the classical Hamiltonian is given by

\begin{equation}
H^{2}=\Lambda (r)\left( M^{2}c^{4}+p^{2}c^{2}+\omega ^{2}(\stackrel{%
\rightarrow }{r}\stackrel{\rightarrow }{p})^{2}\right) .  \label{a.54}
\end{equation}
As in the one-dimensional relativistic oscillator to construct the quantum
mechanical counterpart of (\ref{a.54}), we must respect the ordering
ambiguity of the position- and momentum-operators. Similarly to (\ref{a.5}),
we will be led\ to make the following substitutions:

\begin{equation}
\left\{ 
\begin{array}{c}
x_{i}^{4}p_{i}^{2}\rightarrow -\hbar ^{2}\left( x_{i}^{4}\frac{\partial ^{2}%
}{\partial x_{i}^{2}}+4x_{i}^{3}\frac{\partial }{\partial x_{i}}+\alpha
x_{i}^{2}\right) , \\ 
x_{i}^{2}p_{i}^{2}\rightarrow -\hbar ^{2}\left( x_{i}^{2}\frac{\partial ^{2}%
}{\partial x_{i}^{2}}+2x_{i}\frac{\partial }{\partial x_{i}}+\beta \right) ,
\\ 
x_{i}^{3}p_{i}\rightarrow -i\hbar \left( x_{i}^{3}\frac{\partial }{\partial
x_{i}}+\frac{3}{2}x_{i}^{2}\right) , \\ 
x_{i}p_{i}\rightarrow -i\hbar \left( x_{i}\frac{\partial }{\partial x_{i}}+%
\frac{1}{2}\right) .
\end{array}
\right.  \label{a.55}
\end{equation}
The Green's function $G(\stackrel{\rightarrow }{r^{\prime \prime }}%
,t^{\prime \prime },\stackrel{\rightarrow }{r^{\prime }},t^{\prime })$ for
the problem satisfies the Klein-Gordon equation

\begin{equation}
\left( \Box +U(r)\right) G(\stackrel{\rightarrow }{r}^{\prime \prime
},t^{\prime \prime };\stackrel{\rightarrow }{r}^{\prime },t^{\prime
})=-\frac 1{\hbar ^2c^2}\delta \left( \stackrel{\rightarrow }{r}^{\prime
\prime }-\stackrel{\rightarrow }{r}^{\prime }\right) \delta \left( t^{\prime
\prime }-t^{\prime }\right) .  \label{a.56}
\end{equation}
where

\begin{equation}
\Box =\frac 1{c^2}\frac{\partial ^2}{\partial t^2}-\Lambda (r)\frac
1{r^2}\frac \partial {\partial r}r^2\frac \partial {\partial r}\Lambda
(r)+\Lambda (r)\frac{\widehat{l}^2}{\hbar ^2r^2},  \label{a.57}
\end{equation}
($\widehat{l}^2$ is the square of the orbital angular momentum operator) and 
$U(r)$ is the central potential

\begin{equation}
U(r)=\frac{\hbar ^2\omega ^2}{c^2}\left[ 4\beta -\alpha -\frac{M^2c^4}{\hbar
^2\omega ^2}-8+\left( \alpha +2\beta +\frac 72\right) \Lambda (r)\right] .
\label{a.58}
\end{equation}
It follows that the Green's function $G(\stackrel{\rightarrow }{r^{\prime
\prime }},t^{\prime \prime },\stackrel{\rightarrow }{r^{\prime }},t^{\prime
})$ can be expanded into partial waves\cite{PeakIno} in spherical polar
coordinates

\begin{equation}
G(\stackrel{\rightarrow }{r^{\prime \prime }},t^{\prime \prime },\stackrel{%
\rightarrow }{r^{\prime }},t^{\prime })=\frac 1{r^{\prime \prime }r^{\prime
}}\stackunder{l=0}{\stackrel{\infty }{\sum }}G_l(r^{\prime \prime
},t^{\prime \prime },r^{\prime },t^{\prime })Y_l^{m*}(\theta ^{\prime \prime
},\phi ^{\prime \prime })Y_l^m(\theta ^{\prime },\phi ^{\prime }),
\label{a.59}
\end{equation}
where the radial Green's function, expressed in the Schwinger's integral
representation\cite{Schwinger} , is

\begin{equation}
G_l(r^{\prime \prime },t^{\prime \prime },r^{\prime },t^{\prime })=\frac
1{2i\hbar c^2}\int_0^\infty d\lambda \left\langle r^{\prime \prime
},t^{\prime \prime }\right| \exp \left[ \frac i\hbar \widehat{H}_l\lambda %
\right] \left| r^{\prime },t^{\prime }\right\rangle ,  \label{a.60}
\end{equation}
The integrand in Eq. (\ref{a.60}) is similar to the propagator of an
harmonic oscillator which evolves in the time-like parameter $\lambda $ with
the effective Hamiltonian

\begin{equation}
\widehat{H}_l=\frac 12\left[ -\Lambda (r)\widehat{P}_r^2\Lambda (r)+\frac{%
\widehat{P}_0^2}{c^2}-\hbar ^2l(l+1)\frac{\Lambda (r)}{r^2}+U(r)\right] .
\label{a.61}
\end{equation}

To find the energy eigenvalues $E_{n_{r},l}$ and the wave functions $\Psi
_{n_{r},l}(r)=r^{-1}\Phi _{n_{r},l}(r)$, we may evaluate (\ref{a.60}) by
path integration. The effective Hamiltonian (\ref{a.61}) involves a
centrifugal barrier which possesses a singularity at $r=0$, so that the
discrete form of the expression (\ref{a.60}) is not defined due to a path
collapse. To obtain a tractable and stable path integral, we introduce an
appropriate regulating function ( following Kleinert \cite{Kleinert}) and
write (\ref{a.60}) in the form:

\begin{equation}
G_l(r^{\prime \prime },t^{\prime \prime },r^{\prime },t^{\prime })=\frac
1{2i\hbar c^2}\int_0^\infty dSP_l(r^{\prime \prime },t^{\prime \prime
},r^{\prime },t^{\prime };S),  \label{a.62}
\end{equation}
where the transformed path integral is given in the canonical form by

\begin{eqnarray}
P_l(r^{\prime \prime },t^{\prime \prime },r^{\prime },t^{\prime };S)
&=&f_R(r^{\prime \prime })f_L(r^{\prime })\left\langle r^{\prime \prime
},t^{\prime \prime }\right| \exp \left[ \frac i\hbar Sf_L(r)\widehat{H}%
_lf_R(r)\right] \left| r^{\prime },t^{\prime }\right\rangle  \nonumber \\
&=&f_R(r^{\prime \prime })f_L(r^{\prime })\int Dr(s)Dt(s)\int \frac{%
DP_r(s)DP_0(s)}{(2\pi \hbar )^2}  \nonumber \\
&&\times \exp \left\{ \frac i\hbar \int_0^Sds\left[ -P_r\stackrel{.}{r}+P_0%
\stackrel{.}{t}+f_L(r)H_lf_R(r)\right] \right\}  \nonumber \\
&=&f_R(r^{\prime \prime })f_L(r^{\prime })\stackunder{N\rightarrow \infty }{%
\lim }\stackunder{n=1}{\stackrel{N}{\prod }}\left[ \int dr_ndt_n\right] 
\nonumber \\
&&\times \stackunder{n=1}{\stackrel{N+1}{\prod }}\left[ \int \frac{%
d(P_r)_nd(P_0)_n}{(2\pi \hbar )^2}\right] \exp \left\{ \frac i\hbar 
\stackunder{n=1}{\stackrel{N+1}{\sum }}A_1^{\varepsilon _s}\right\} ,
\label{a.63}
\end{eqnarray}
with the short-time action

\begin{eqnarray}
A_1^{\varepsilon _s} &=&-(P_r)_n\triangle r_n+(P_0)_n\triangle t_n+\frac{%
\varepsilon _s}2f_L(r_n)\left[ -\Lambda (r_n)\Lambda (r_{n-1})(P_r)_n^2+%
\frac{(P_0)_n^2}{c^2}\right.  \nonumber \\
&&\left. -\hbar ^2l(l+1)\frac{\Lambda (r_n)}{r_n^2}+U(r_n)\right]
f_R(r_{n-1}),  \label{a.64}
\end{eqnarray}
and

\begin{equation}
\varepsilon _{s}=\frac{S}{N+1}=\triangle s_{n}=\frac{\triangle \tau _{n}}{%
f_{L}(r_{n})f_{R}(r_{n-1})};\qquad \triangle \tau _{n}=\varepsilon _{\tau }=%
\frac{\lambda }{N+1}.  \label{a.65}
\end{equation}
The regulating function is defined as \cite{Kleinert}

\begin{equation}
f(r)=f_L(r)f_R(r)=f^{1-\lambda ^{\prime }}(r)f^{\lambda ^{\prime }}(r).
\label{a.66}
\end{equation}

As in the $(1+1)$-dimensional case, by doing successively the $t_n$ and $%
(P_0)_n$ integrations we arrive at

\begin{equation}
P_l(r^{\prime \prime },t^{\prime \prime },r^{\prime },t^{\prime };S)=\frac
1{2\pi \hbar }\int_{-\infty }^{+\infty }dE\exp \left[ -\frac i\hbar
E(t^{\prime \prime }-t^{\prime })\right] P_l(r^{\prime \prime },r^{\prime
};S),  \label{a.67}
\end{equation}
where the invariant kernel $P_l(r^{\prime \prime },r^{\prime };S)$ under the
change $E\rightarrow -E$ is given by

\begin{eqnarray}
P_{l}(r^{\prime \prime },r^{\prime };S) &=&f_{R}(r^{\prime \prime
})f_{L}(r^{\prime })\stackunder{N\rightarrow \infty }{\lim }\stackunder{n=1}{%
\stackrel{N}{\prod }}\left[ \int dr_{n}\right]  \nonumber \\
&&\times \stackunder{n=1}{\stackrel{N+1}{\prod }}\left[ \int \frac{%
d(P_{r})_{n}}{(2\pi \hbar )}\right] \exp \left\{ \frac{i}{\hbar }\stackunder{%
n=1}{\stackrel{N+1}{\sum }}A_{2}^{\varepsilon _{s}}\right\} ,  \label{a.68}
\end{eqnarray}
with

\begin{eqnarray}
A_2^{\varepsilon _s} &=&-(P_r)_n\triangle r_n+\frac{\varepsilon _s}2f_L(r_n)%
\left[ -\Lambda (r_n)\Lambda (r_{n-1})(P_r)_n^2+\frac{E^2}{c^2}\right. 
\nonumber \\
&&\left. -\hbar ^2l(l+1)\frac{\Lambda (r_n)}{r_n^2}+U(r_n)\right]
f_R(r_{n-1}).  \label{a.69}
\end{eqnarray}
Substituting (\ref{a.67}) into (\ref{a.62}), we observe that the t-dependent
term does not contain the variable $S$. Therefore, we can rewrite the
partial Green's function (\ref{a.62}) in the form:

\begin{equation}
G_l(r^{\prime \prime },t^{\prime \prime },r^{\prime },t^{\prime })=\frac
1{2\pi \hbar }\int_{-\infty }^{+\infty }dE\exp \left[ -\frac i\hbar
E(t^{\prime \prime }-t^{\prime })\right] G_l(r^{\prime \prime },r^{\prime }),
\label{a.70}
\end{equation}
with

\begin{equation}
G_l(r^{\prime \prime },r^{\prime })=\frac 1{2i\hbar c^2}\int_0^\infty
dSP_l(r^{\prime \prime },r^{\prime };S).  \label{a.71}
\end{equation}

The path integration of the kernel $P_l(r^{\prime \prime },r^{\prime };S)$
can be performed for any splitting parameter $\lambda ^{\prime }$. However,
to simplify the calculation, we prefer to work with the mid-point
prescription by taking $\lambda ^{\prime }=\frac 12$. This can be justified
by the fact that the final result is independent of this parameter. Then, by
integrating with respect to the variables $(P_r)_n$, we find

\begin{eqnarray}
P_{l}(r^{\prime \prime },r^{\prime };S) &=&\frac{\left[ f(r^{\prime
})f(r^{\prime \prime })\right] ^{\frac{1}{4}}}{\sqrt{\Lambda (r^{\prime
})\Lambda (r^{\prime \prime })}}\stackunder{N\rightarrow \infty }{\lim }%
\stackunder{n=1}{\stackrel{N+1}{\prod }}\sqrt{\frac{1}{2i\pi \hbar
\varepsilon _{s}}}  \nonumber \\
&&\times \stackunder{n=1}{\stackrel{N}{\prod }}\left[ \int \frac{dr_{n}}{%
\Lambda (r_{n})\sqrt{f(r_{n})}}\right] \exp \left\{ \frac{i}{\hbar }%
\stackunder{n=1}{\stackrel{N+1}{\sum }}A_{3}^{\varepsilon _{s}}\right\}
\label{a.72}
\end{eqnarray}
with the short-time action in configuration space

\begin{eqnarray}
A_3^{\varepsilon _s} &=&\frac{\triangle r_n^2}{2\varepsilon _s\Lambda
(r_n)\Lambda (r_{n-1})\sqrt{f(r_n)f(r_{n-1})}}+\frac{\varepsilon _s}2f(r_n)%
\left[ \frac{E^2}{c^2}\right.  \nonumber \\
&&\left. -\hbar ^2l(l+1)\frac{\Lambda (r_n)}{r_n^2}+U(r_n)\right] .
\label{a.73}
\end{eqnarray}

We now use the following space transformation: $r\rightarrow u,$ $r\in \left[
0,\infty \right[ ,$ $u\in \left] -\infty ,\infty \right[ $ defined by

\begin{equation}
r=\frac{c}{\omega }e^{u}.  \label{a.74}
\end{equation}
The appropriate regulating function is then defined by

\begin{equation}
f(r(u))=\frac{c^2}{4\omega ^2\cosh ^2u}.  \label{a.75}
\end{equation}

By taking into account all the quantum corrections arising, of course, from
the transformations (\ref{a.74}) and (\ref{a.75}), the Green's function (\ref
{a.71}) can straightforward be written as follows:

\begin{equation}
G_l(r^{\prime \prime },r^{\prime })=\frac 1{4i\hbar \omega c\sqrt{\Lambda
(r^{\prime \prime })\Lambda (r^{\prime })\cosh u^{\prime \prime }\cosh
u^{\prime }}}\int_0^\infty dSP_l(u^{\prime \prime },u^{\prime };S),
\label{a.76}
\end{equation}
with

\begin{eqnarray}
P_l(u^{\prime \prime },u^{\prime };S) &=&\int Du(s)\exp \left\{ \frac i\hbar
\int_0^S\left[ \frac{\stackrel{.}{u}^2}2-\frac{\hbar ^2}4\left( \nu
^2+k^2\right. \right. \right.  \nonumber \\
&&\left. \left. \left. +\left( \nu ^2-k^2\right) \tanh u\right) +\frac{\hbar
^2}8\frac{\frac{E^2}{\hbar ^2\omega ^2}+4\beta -\alpha -2}{\cosh ^2u}\right]
\right\} ,  \label{a.77}
\end{eqnarray}
where $\nu ^2=N^2-2\beta -\alpha +\frac{11}4,$ $k=l+\frac 12$ and $%
N=Mc^2/\hbar \omega .$

This kernel is formally identical with that of the general Rosen-Morse ( or
general modified P\"oschl-Teller ) potential studied recently \cite
{JunIno,BohJun,Grosc,KleiMus} . The Green's function associated to this
potential is

\begin{equation}
G(u^{\prime \prime },u^{\prime };E_{RM^{\prime }})=\int_{0}^{\infty
}dSP_{l}(u^{\prime \prime },u^{\prime };S).  \label{a.78}
\end{equation}
As is shown by Kleinert \cite{Kleinert} , the Green's function of the
general Rosen-Morse potential is related to the fixed-energy amplitude for
the mass point subjected to an angular barrier near the surface of a sphere
in $D=4$ dimensions by

\begin{eqnarray}
G(u^{\prime \prime },u^{\prime };E_{RM^{\prime }}) &=&\frac{1}{\sqrt{\sin
\theta ^{\prime \prime }\sin \theta ^{\prime }}}G(\theta ^{\prime \prime
},\theta ^{\prime };E_{PT^{\prime }})  \nonumber \\
&=&-\frac{i}{\hbar }\frac{\Gamma (M_{1}-L_{E})\Gamma (L_{E}+M_{1}+1)}{\Gamma
(M_{1}+M_{2}+1)\Gamma (M_{1}-M_{2}+1)}  \nonumber \\
&&\times \left( \frac{1+\tanh u^{\prime }}{2}\right)
^{(M_{1}-M_{2})/2}\left( \frac{1-\tanh u^{\prime }}{2}\right)
^{(M_{1}+M_{2})/2}  \nonumber \\
&&\times \left( \frac{1-\tanh u^{\prime \prime }}{2}\right)
^{(M_{1}+M_{2})/2}\left( \frac{1+\tanh u^{\prime \prime }}{2}\right)
^{(M_{1}-M_{2})/2}  \nonumber \\
&&\times F\left( M_{1}\!-\!L_{E},L_{E}\!+\!M_{1}\!+\!1;M_{1}\!-\!M_{2}\!+\!1;%
\frac{1+\tanh u^{\prime }}{2}\right)  \nonumber \\
&&\times F\left( M_{1}\!-\!L_{E},L_{E}\!+\!M_{1}\!+\!1;M_{1}\!+\!M_{2}\!+\!1;%
\frac{1-\tanh u^{\prime \prime }}{2}\right) ,  \nonumber \\
&&  \label{a.79}
\end{eqnarray}
with $\tanh u=-\cos \theta ,\theta \in \left( 0,\pi \right) ,u\in \left]
-\infty ,+\infty \right[ $ and $u^{\prime \prime }>u^{\prime }.$ Here, the
mass point is taken equal to unity. In addition, we set

\begin{equation}
\left\{ 
\begin{array}{c}
L_E=-\frac 12+\left( \frac 1{16}+\frac{2E_{PT^{\prime }}}\hbar \right)
^{\frac 12}, \\ 
E_{PT^{\prime }}=\frac{\hbar ^2}8\left( \frac{E^2}{\hbar ^2\omega ^2}+4\beta
-\alpha -\frac 54\right) ,
\end{array}
\right.  \label{a.80}
\end{equation}
and if we choose 
\begin{equation}
\left\{ 
\begin{array}{c}
M_1=\frac 12\left( \sqrt{N^2-2\beta -\alpha +\frac{11}4}+l+\frac 12\right) ,
\\ 
M_2=\frac 12\left( \sqrt{N^2-2\beta -\alpha +\frac{11}4}-l-\frac 12\right) ,
\end{array}
\right.  \label{a.81}
\end{equation}
the boundary conditions for the wave functions appearing in (\ref{a.79})
will be satisfied.

The equivalence between the relativistic harmonic oscillator interaction in $%
(3+1)$ Minkowski space-time and a free relativistic particle in CAdS is
characterized by the following restriction on the parameters $\alpha $ and $%
\beta $:

\begin{equation}
\alpha =8\xi ,\qquad \beta =2\xi +\frac{1}{4}.  \label{a.82}
\end{equation}
Inserting (\ref{a.79}) into (\ref{a.76}), we get, for the radial Green's
function, the closed form:

\begin{eqnarray}
G_{l}(r^{\prime \prime },r^{\prime }) &=&-\frac{\Gamma (M_{1}-L_{E})\Gamma
(L_{E}+M_{1}+1)}{4\hbar ^{2}\omega c\Gamma (M_{1}+M_{2}+1)\Gamma
(M_{1}-M_{2}+1)}  \nonumber \\
&&\times \left( \Lambda (r^{\prime \prime })\Lambda (r^{\prime })\cosh
u^{\prime \prime }\cosh u^{\prime }\right) ^{-\frac{1}{2}}  \nonumber \\
&&\times \left( \frac{1+\tanh u^{\prime }}{2}\right)
^{(M_{1}-M_{2})/2}\left( \frac{1-\tanh u^{\prime }}{2}\right)
^{(M_{1}+M_{2})/2}  \nonumber \\
&&\times \left( \frac{1-\tanh u^{\prime \prime }}{2}\right)
^{(M_{1}+M_{2})/2}\left( \frac{1+\tanh u^{\prime \prime }}{2}\right)
^{(M_{1}-M_{2})/2}  \nonumber \\
&&\times F\left( M_{1}-L_{E},L_{E}+M_{1}+1;M_{1}-M_{2}+1;\frac{1+\tanh
u^{\prime }}{2}\right)  \nonumber \\
&&\times F\left( M_{1}-L_{E},L_{E}+M_{1}+1;M_{1}+M_{2}+1;\frac{1-\tanh
u^{\prime \prime }}{2}\right) .  \nonumber \\
&&  \label{a.83}
\end{eqnarray}
The poles of (\ref{a.83}) are all contained in the first $\Gamma $ function
in the numerator,

\begin{equation}
M_1-L_E=-n_r.  \label{a.84}
\end{equation}
Converting this into energy by using Eqs. (\ref{a.80}), (\ref{a.81}) and (%
\ref{a.84}) yields

\begin{equation}
E_{n_{r},l}=\beta _{n_{r},l}\hbar \omega ,  \label{a.85}
\end{equation}
with 
\begin{equation}
\beta _{n_{r},l}=2n_{r}+l+\frac{1}{2}\sqrt{9+4N^{2}-48\xi }+\frac{3}{2},
\label{a.86}
\end{equation}
where $n_{r}$ is the radial quantum number and $l$ the angular momentum.
Here, the parameter $\xi $ is subject to the condition $\xi <\frac{1}{48}%
(9+4N^{2})$.

In the non-relativistic approximation

\begin{equation}
E_{n_r,l}\stackunder{c\rightarrow \infty }{\rightarrow }\left( 2n_r+l+\frac
32\right) \hbar \omega +Mc^2,  \label{a.87}
\end{equation}
where the first term represents the well-known energy spectrum of the
three-dimensional non-relativistic harmonic oscillator.

As in the one-dimensional case, the radial wave functions can be found by
approximation near the poles $M_1-L_E\approx -n_r$:

\begin{equation}
\Gamma (M_{1}-L_{E})\approx \frac{(-1)^{n_{r}}}{n_{r}!}\frac{1}{%
M_{1}-L_{E}+n_{r}}=\frac{(-1)^{n_{r}+1}}{n_{r}!}\frac{4\hbar ^{2}\omega
^{2}\beta _{n_{r},l}}{E^{2}-E_{n_{r},l}^{2}}.  \label{a.88}
\end{equation}
Using this behavior and taking into consideration the Gauss transformation
formula (see Eq. (9.131.2) , p. 1043 in Ref.\cite{GradRyz} )

\begin{eqnarray}
F(a,b,c;z) &=&\frac{\Gamma (c)\Gamma (c\!-\!a\!-\!b)}{\Gamma (c\!-\!a)\Gamma
(c\!-\!b)}F(a,b,a\!+\!b\!-\!c\!+\!1;1\!-\!z)+\frac{\Gamma (c)\Gamma
(a\!+\!b\!-\!c)}{\Gamma (a)\Gamma (b)}  \nonumber \\
&&\times (1\!-\!z)^{c-a-b}F(c\!-\!a,c\!-\!b,c\!-\!a\!-\!b\!+\!1;1\!-\!z),
\label{a.89}
\end{eqnarray}
knowing that the second term of this latter is null because the Euler
function $\Gamma (a)$ is infinite ( $a=-n_{r}\leq 0$ ), we can write Eq. (%
\ref{a.83}) as:

\begin{equation}
G_{l}(r^{\prime \prime },r^{\prime })=\stackunder{n_{r}=0}{\stackrel{\infty 
}{\sum }}\frac{\Phi _{n_{r},l}(r^{\prime \prime })\Phi _{n_{r},l}^{\ast
}(r^{\prime })}{E^{2}-E_{n_{r},l}^{2}},  \label{a.90}
\end{equation}
where 
\begin{eqnarray}
\Phi _{n_{r},l}(r) &=&\left[ \frac{2\beta _{n_{r},l}\Gamma \left( \beta
_{n_{r},l}-n_{r}\right) \Gamma \left( n_{r}+l+\frac{3}{2}\right) }{%
n_{r}!\Gamma \left( \beta _{n_{r},l}-n_{r}-\frac{1}{2}\right) }\right] ^{%
\frac{1}{2}}\frac{1}{\Gamma \left( l+\frac{3}{2}\right) }\left( \frac{\omega 
}{c}\right) ^{l+\frac{3}{2}}r^{l}  \nonumber \\
&&\!\!\!\!\times \left( 1\!+\!\frac{\omega ^{2}}{c^{2}}r^{2}\right)
^{n_{r}-\beta _{n_{r},l}/2}F\left( \!\!-n_{r},\beta _{n_{r},l}-n_{r};l+\frac{%
3}{2};\frac{\frac{\omega ^{2}}{c^{2}}r^{2}}{1+\frac{\omega ^{2}}{c^{2}}r^{2}}%
\!\!\right)  \label{a.91}
\end{eqnarray}
are the radial wave functions.

By substituting (see Eq. (9.131.1), p. 1043 in Ref.\cite{GradRyz} )

\begin{equation}
F(\alpha ,\beta ;\gamma ;z)=(1-z)^{-\alpha }F(\alpha ,\gamma -\beta ;\gamma
;\frac z{z-1})  \label{a.92}
\end{equation}
into (\ref{a.91}) and using the connecting formula (see Eq. (8.962.1), p.
1036 in Ref.\cite{GradRyz} )

\begin{equation}
P_n^{(\alpha ,\beta )}(x)=\frac{\Gamma (n+\alpha +1)}{n!\Gamma (\alpha +1)}%
F\left( -n,n+\alpha +\beta +1;\alpha +1;\frac{1-x}2\right) ,  \label{a.93}
\end{equation}
we can also express (\ref{a.91}) in the form

\begin{eqnarray}
\Phi _{n_r,l}(r) &=&\left[ \frac{2\beta _{n_r,l}n_r!\Gamma \left( \beta
_{n_r,l}-n_r\right) }{\Gamma \left( n_r+\gamma +1\right) \Gamma \left(
n_r+l+\frac 32\right) }\right] ^{\frac 12}\left( \frac \omega c\right)
^{l+\frac 32}r^l  \nonumber \\
&&\times \left( 1+\frac{\omega ^2}{c^2}r^2\right) ^{-\frac 12\beta
_{n_r,l}}P_{n_r}^{\left( l+\frac 12,-\beta _{n_r,l}\right) }\left( 1+2\frac{%
\omega ^2}{c^2}r^2\right) .  \label{a.94}
\end{eqnarray}

By using the following limiting relations:

\begin{equation}
\left\{ 
\begin{array}{c}
\stackunder{\gamma \rightarrow \infty }{\lim }\Gamma (n+\gamma +1)=%
\stackunder{\gamma \rightarrow \infty }{\lim }\gamma ^{n+1}\Gamma (\gamma ),
\\ 
\stackunder{\gamma \rightarrow \infty }{\lim }\left( 1+\frac{\omega ^2}{c^2}%
r^2\right) ^{-\frac 12\left( l+\gamma +\frac 32\right) }=e^{-\frac{M\omega }{%
2\hbar }r^2}, \\ 
\stackunder{\gamma \rightarrow \infty }{\lim }F\left( \alpha ,\gamma ;\nu
;\frac z\gamma \right) =F(\alpha ,\nu ;z),
\end{array}
\right.  \label{a.95}
\end{equation}
we obtain the well-known radial wave functions of the non-relativistic
harmonic oscillator

\begin{eqnarray}
\Phi _{n_r,l}(r) &=&\left[ \frac{2\Gamma \left( n_r+l+\frac 32\right) }{n_r!}%
\right] ^{\frac 12}\left( \frac{M\omega }\hbar \right) ^{\frac 12\left(
l+\frac 32\right) }\frac{r^l}{\Gamma \left( l+\frac 32\right) }  \nonumber \\
&&\times \exp \left[ -\frac{M\omega }{2\hbar }r^2\right] F\left(
-n_r,l+\frac 32;\frac{M\omega }\hbar r^2\right) .  \label{a.96}
\end{eqnarray}

\section{Conclusion}

In this paper we have dealt with special relativistic harmonic oscillators
in $(1+1)-$ and $(3+1)-$dimensional Minkowski space-time modeled by a free
relativistic particle in the universal covering space-time of the anti-de
Sitter space-time. The explicit path integral solution, as presented above,
provides a valuable alternative way to the one obtained through the
Klein-Gordon equation. After formulating the problem in terms of symmetric
and general Rosen-Morse potentials for the one- and three-dimensional
relativistic oscillators, respectively and by imposing a restriction on the
parameters $\alpha $ and $\beta $ in such a way that the system under
consideration is equivalent to a free relativistic particle in CAdS, the
Green's functions are obtained in a closed form. The energy spectrum and the
properly normalized wave functions are extracted from the poles and the
residues at the poles of the Green's function, respectively. In the
flat-space limit $(R\rightarrow 0)$, that is to say in the non-relativistic
approximation $(c\rightarrow \infty )$, the usual harmonic oscillator
spectrum and the corresponding normalized wave functions are regained.

\end{document}